\documentclass[pra,twocolumn,superscriptaddress,amsmath,amssymb,longbibliography]{revtex4-1}
\usepackage{hyperref}
\usepackage{graphicx}
\usepackage{array}
\usepackage{amsmath,amsfonts,amssymb}
\usepackage[ansinew]{inputenc}
\usepackage{color}
\usepackage{calc}
\usepackage{ulem}

\newcommand{\Figref}[1]{Fig.~\ref{#1}}

\begin{document}
\title{Tip-induced excitonic luminescence nanoscopy of an atomically-resolved van der Waals heterostructure}

\author{Luis E. Parra L\'opez}
\affiliation{Universit\'e de Strasbourg, CNRS, Institut de Physique et Chimie des Mat\'eriaux de Strasbourg, UMR 7504, F-67000 Strasbourg, France}

\author{Anna Ros\l awska}
\affiliation{Universit\'e de Strasbourg, CNRS, Institut de Physique et Chimie des Mat\'eriaux de Strasbourg, UMR 7504, F-67000 Strasbourg, France}

\author{Fabrice Scheurer}
\affiliation{Universit\'e de Strasbourg, CNRS, Institut de Physique et Chimie des Mat\'eriaux de Strasbourg, UMR 7504, F-67000 Strasbourg, France}

\author{St\'ephane Berciaud}
\email{stephane.berciaud@ipcms.unistra.fr}
\affiliation{Universit\'e de Strasbourg, CNRS, Institut de Physique et Chimie des Mat\'eriaux de Strasbourg, UMR 7504, F-67000 Strasbourg, France}

\author{Guillaume Schull}
\email{guillaume.schull@ipcms.unistra.fr}
\affiliation{Universit\'e de Strasbourg, CNRS, Institut de Physique et Chimie des Mat\'eriaux de Strasbourg, UMR 7504, F-67000 Strasbourg, France}

\begin{abstract}

Low-temperature scanning tunneling microscopy is used to probe, with atomic-scale spatial resolution, the intrinsic luminescence of a van der Waals heterostructure, made of a transition metal dichalcogenide monolayer stacked onto a few-layer graphene flake supported by an Au(111) substrate. Sharp emission lines arising from neutral, charged and localised excitons are reported. Their intensities and emission energies vary as a function of the nanoscale environment of the van der Waals heterostructure, explaining the variability of the emission properties observed with diffraction-limited approaches. Our work paves the way towards understanding and control of optoelectronic phenomena in moir\'e superlattices with atomic-scale resolution.

\end{abstract}


\maketitle

\section*{Introduction}

Van der Waals (vdW) heterostructures made from stacks of two-dimensional (2D) materials, in particular semiconducting transition metal dichalcogenides (TMDs), are ideal systems for studying fundamental phenomena related to the confinement of tightly bound excitons~\cite{Wangreviewexcitons2018} and their exploitation in atomically thin optoelectronic devices~\cite{Mak2016,Wilson2021}. The rich physics of excitons, trions (i.e., charged excitons) and more complex many-body states in TMDs and related vdW heterostructures has been addressed through optical spectroscopy~\cite{Wangreviewexcitons2018,Shree2021}, in particular by recording their micro-photoluminescence ($\mu$PL) characteristics, with a diffraction-limited spatial resolution of typically $\lesssim 1~ \rm{\mu m}$. These far-field optical studies have highlighted significant spatial variations in the optical response of a given sample, due to inhomogeneities of the nanoscale environment, including strain gradients~\cite{Harats2020}, dielectric disorder~\cite{Raja2019}, as well as to localised defects~\cite{Tongay2013a,Chow2015} and dopants~\cite{Tongay2013b}. Taking advantage of these subtle structure/property relationships, periodic nanoscale moir\'e super-potentials resulting from the controlled rotational mismatch between stacked 2D layers have recently been used to tailor exciton physics, with potential outcomes for quantum simulation and quantum technologies~\cite{Tang2020,Li2021,Wilson2021,Kennes2021}. Understanding of such emergent phenomena requires addressing excitons and their local-environment with a spatial resolution below the moir\'e wavelength ($\lesssim 10~\rm{nm}$) and the exciton Bohr radius ($\sim 1~\rm{nm}$)\cite{Yu2017,Huang2022}, i.e., two to three orders of magnitude below the optical diffraction limit.\\
Recently, attempts have been made to address excitons in TMDs with nanoscale resolution~\cite{Park2018,Darlington2020,Bonnet2021,Pommier2019,pechou2020plasmonic,Pena2020,Zhang2022}. In particular, tip-enhanced PL studies have provided original ways to control the emission characteristics of TMD monolayers~\cite{Park2018} as well as to image nanoscale strain gradients~\cite{Darlington2020} in ambient conditions with a resolution of $\approx$ 15~nm.  In parallel, atomically-resolved luminescence has been reported on single molecules using scanning tunneling microscopy-induced luminescence (STML)\cite{Qiu2003,Wu2008,Zhang2016,Doppagne2017,Zhang2017sps,Doppagne2018,Kimura2019}, a method that was recently applied to investigate excitonic emission from TMDs in ambient air~\cite{Pommier2019,pechou2020plasmonic,Pena2020}. Under these conditions, atomic-scale resolution could not be attained, probably because of atmospheric contamination and lack of mechanical and thermal stability. Attempts to address the excitonic properties of TMDs with STM in ultra-high vacuum (UHV) at cryogenic temperatures have been reported~\cite{Krane2016,Schuler2020}, but the luminescence signals were dominated by inelastic transitions between electrodes and TMD states, whereas radiative recombination of excitons and trions was likely quenched by the strong interaction with the supporting metallic substrate~\cite{Velicky2018}. Approaches allowing investigating excitonic emission from vdW heterostructures at the atomic scale are highly desirable, but still lacking.\\
In this article, we take up this challenge and demonstrate intrinsic excitonic luminescence from a two-dimensional semiconductor with the nanoscale-resolution provided by low-temperature STML. We show that a vdW heterostructrure based on a TMD monolayer, decoupled from an Au(111) crystal by a few-layer graphene flake (FLG) allows preserving its luminescence while ensuring optimal STM imaging of surface atoms and moir\'e superlattices. Our results provide insights into the mechanisms leading to STM-induced luminescence in vdW heterostructures and further establish STM-based methods as a unique tool to correlate the optical response of low-dimensional systems to their nano- and atomic-scale environment.

\noindent

\section*{Results and discussion}

\begin{figure}[!th]
  \includegraphics[width=0.75\linewidth]{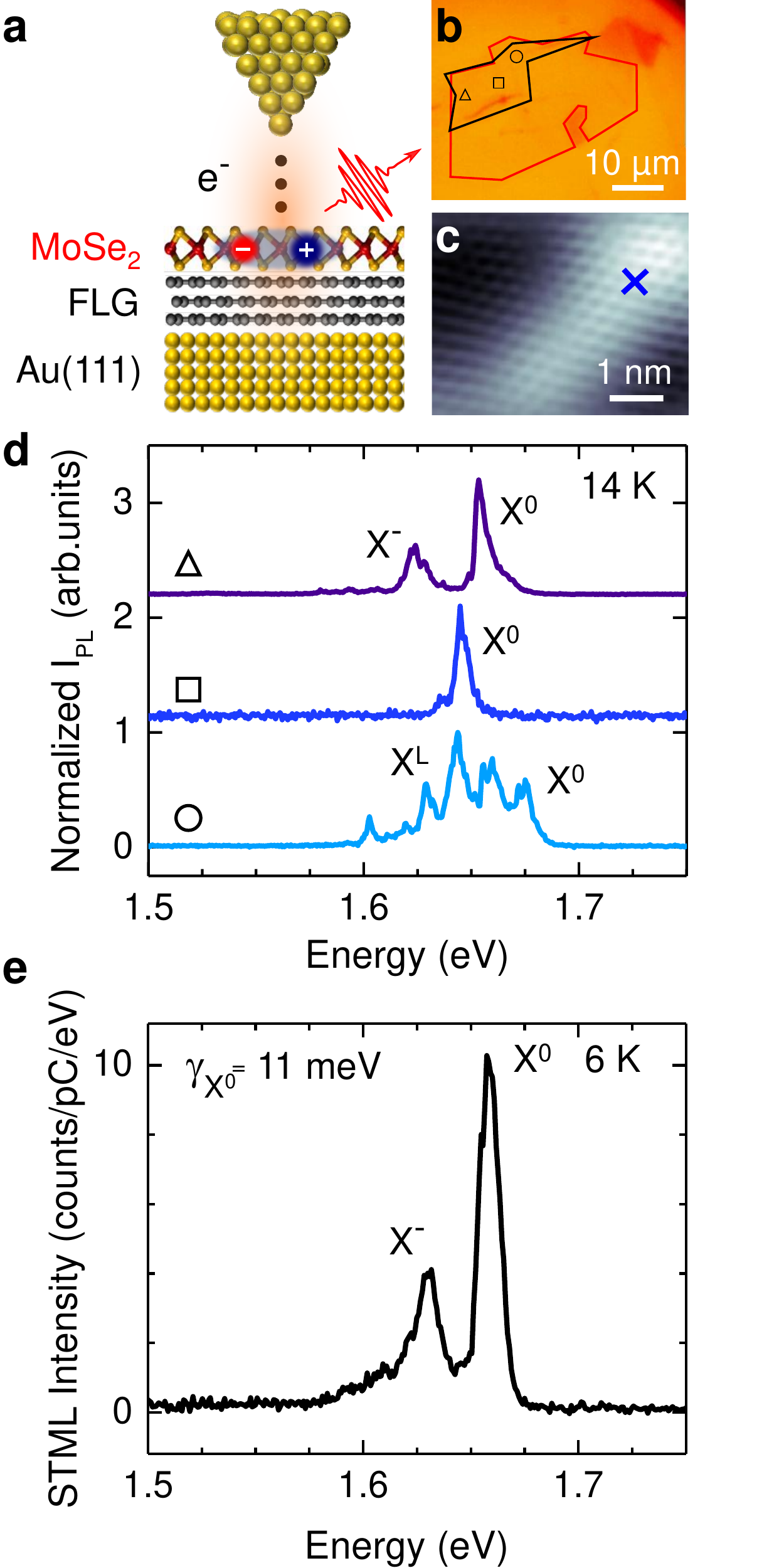}
  \caption{\label{fig1}\textbf{STM-induced luminescence of a MoSe$_2$/FLG/Au(111) heterostructure.} \textbf{a.} Artistic view of the STML experiment. The red curved arrow represents a photon emitted following the radiative recombination of excitons. ~\textbf{b.} Optical microscopy image of the sample. The monolayer of MoSe$_2$ is highlighted in red and the FLG in black, while the underlying Au(111) covers the rest of the image.~ \textbf{c.} Atomically-resolved constant-current STM image ($V = -1.3~$V and $I = 10~$pA) of the heterostructure surface. \textbf{d.} Series of normalised $\mu$PL spectra acquired on the different points of the heterostructure shown in \textbf{b.} The spectra were acquired on a home-built $\mu$PL setup at 14~K under continuous wave laser-excitation centered at 532~nm and under an incident laser intensity of $\sim 30 ~ \mu$W/$\mu \rm{m}^2$. \textbf{e.} STML spectrum recorded on MoSe$_2$/FLG/Au(111) in the cross-marked area on \textbf{c}, with $V = -2.8~$V, $I = 90~\rm{pA}$. The labels $\rm X^0$, $\rm X^-$ and $\rm X^{\rm L}$ in \textbf{d} and \textbf{e} denote the neutral, negatively charged and localised excitons, respectively. The FWHM of the $\rm X^0$ line (denoted $\gamma_{\rm{X}^0}$) is shown in the upper left corner of \textbf{e}.}
\end{figure}

Our measurements were performed on a vdW heterostructure made of a molybdenum diselenide (MoSe$_2$ ) monolayer stacked on top of a few layer ($\sim 3-5$) graphene flake (FLG) deposited onto an Au(111) substrate (see Methods and Supplementary Section S1 for more details). Figure \ref{fig1}a, b and c  show a schematic  representation of the STM-induced luminescence (STML) experiment, as well as an optical image of the MoSe$_2$/FLG/Au(111) heterostructure and an atomically-resolved constant current STM image, respectively. As further discussed below, a STM silver (Ag) tip is used to locally generate excitonic electroluminescence from the MoSe$_2$ monolayer. Here, akin to hexagonal boron nitride (hBN)~\cite{Dean2010,Cadiz2017}, the FLG provides a smooth substrate for the TMD, while ensuring efficient electrical conduction. This configuration also makes use of the plasmonic properties of the Ag-tip Au-substrate junction  (Fig.~\ref{fig1}a), which is known to enhance radiative recombination in STML experiments~\cite{Roslawska2022}, leading to an optimal configuration to detect excitonic luminescence.

\begin{figure*}[!th]
  \includegraphics[width=0.7\linewidth]{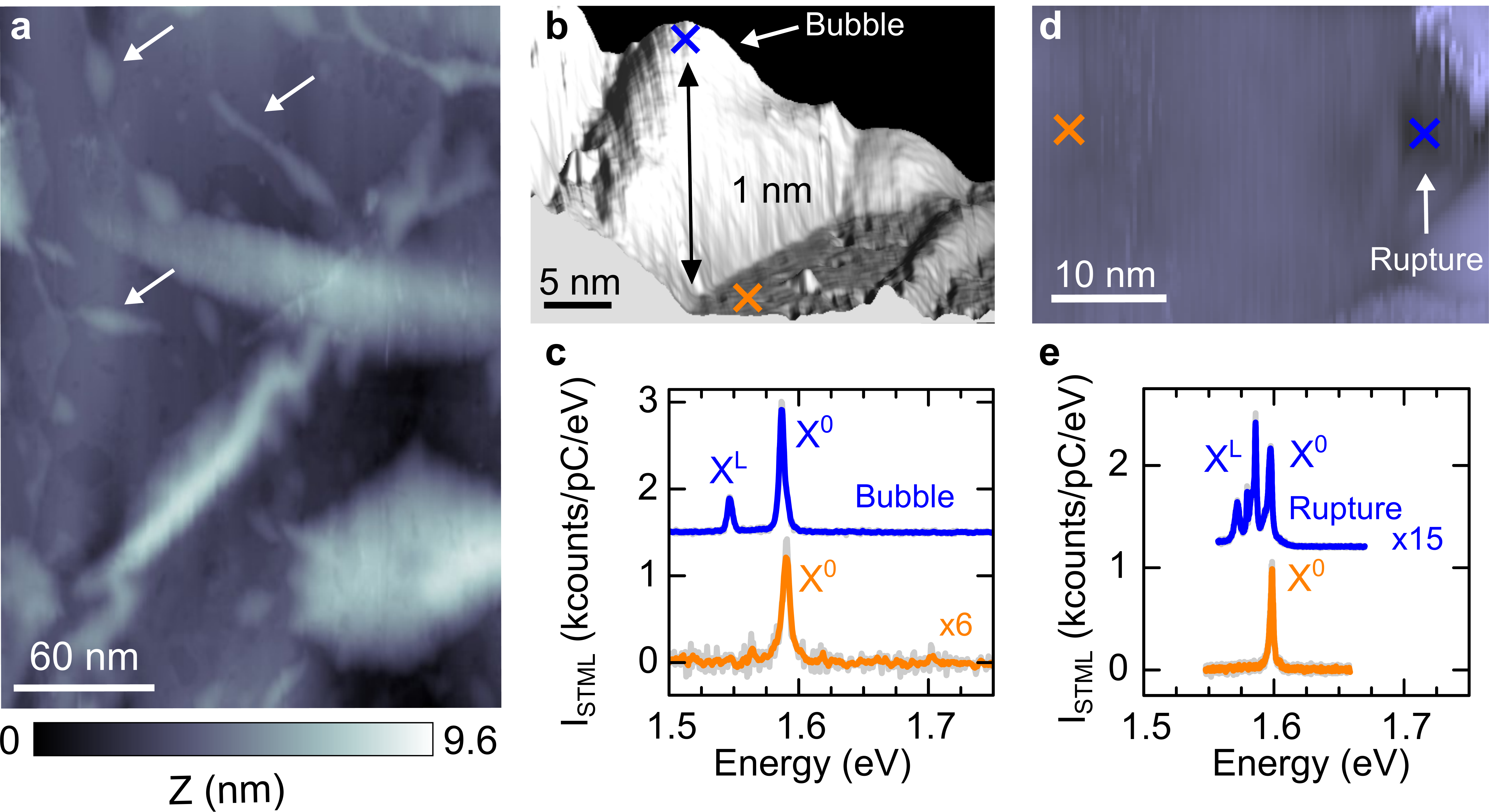}
  \caption{\label{fig2} \textbf{Spatially-resolved STML in an inhomogeneous nanoscale landscape.}~\textbf{a.} Constant-current STM image ($V = -2.2~$V, $I = 4~$pA) of the MoSe$_2$/FLG surface. During the fabrication process the heterostructure conforms to the substrate underneath, resulting in nanoscale folds, ripples and bubbles (white arrows).\textbf{b.} Constant-current STM image ($V = -2.1~$V, $I = 4~$pA)  of a nano-bubble on the heterostructure surface. \textbf{c.} Normalised STML spectra acquired at $V = -3.2~$V on the marked areas of image \textbf{b} with 4 pA (blue spectrum) and 10 pA (orange spectrum). \textbf{d} Constant-current STM image ($V = -1.85~$V, $I = 5~$pA) of an area displaying a rupture on the top layer of the heterostructure. Two areas corresponding to the rupture (blue cross) and the flat area (orange cross) are shown. \textbf{e.} Normalised STML spectra acquired under $V = -1.85~$V, $I = 5~$ pA on the marked areas of image \textbf{b} corresponding to rupture (blue spectrum) or flat areas (orange spectrum). The spectra in \textbf{c} and \textbf{e} are vertically offset for clarity.} 
\end{figure*}

\begin{figure*}[!th]
  \includegraphics[width=0.55\linewidth]{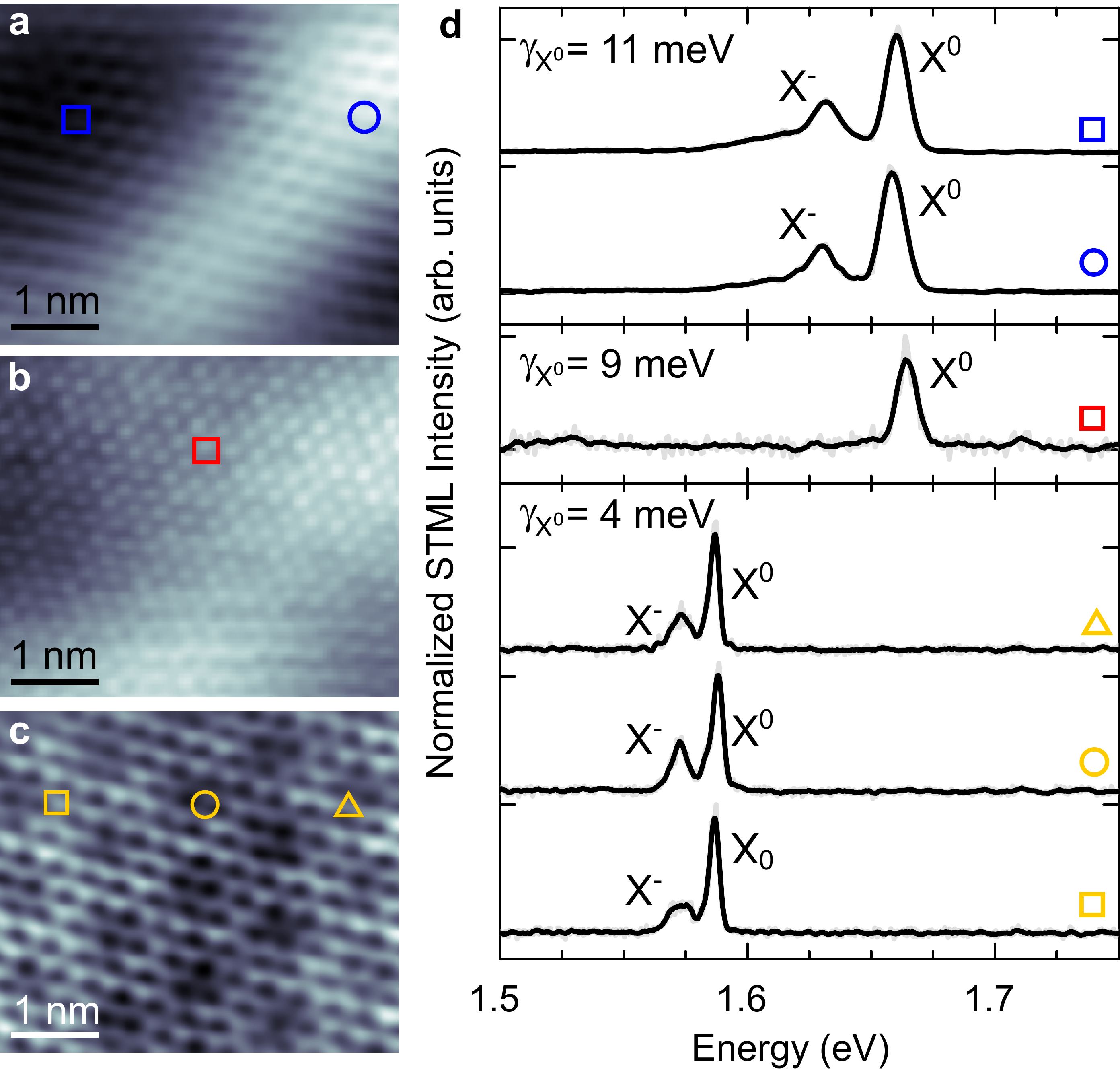}
  \caption{\label{fig3} \textbf{STML on atomically-resolved areas.}~ \textbf{a},\textbf{b} and \textbf{c} are atomically-resolved constant-current STM images of three different areas of the MoSe$_2$/FLG surface ($V = -1.3~$ V, $I = 10~$ pA for \textbf{a}; $V = -1.4~$V, $I = 4.1~$pA for \textbf{b}; $V = -1.4~$V, $I = 6~$pA for \textbf{c}). Each region corresponds to different interface qualities of the heterostructure. \textbf{d.} STML spectra recorded on the different regions at locations marked by symbols ($V = -2.8~$V, $I = 90~$pA; for \textbf{a}; $V = -2.8~$V, $I = 100~$pA for \textbf{b}; $V = - 2.9~$V, $I = 90~$pA for \textbf{c}).} 
\end{figure*}

 We first report on low temperature $\mu$PL measurements recorded \textit{ex situ} using a laser beam of $\approx 1~\mu \rm m$ in diameter at various close-lying spots on the sample, indicated in Fig.~\ref{fig1}b. The strikingly different $\mu$PL spectra shown in Fig.~\ref{fig1}d suggest sizeable inhomogeneities at the sub-micrometre scale. These inhomogeneities are likely due to the necessary thermal annealing  step performed before introducing the sample into the STM chamber (see Methods) and are further resolved using STML, as discussed in Fig.~\ref{fig2} and ~\ref{fig3}.

 Three types of low temperature $\mu$PL spectra are identified on the MoSe$_2$/FLG/Au(111) region shown in Fig.~\ref{fig1}b (see Supplementary Section 2 for a detailed analysis). First, we observe spectra dominated by a high-energy emission line near 1.65~eV, followed by a lower intensity feature, situated about 30~meV below the main line (top trace in Fig.~\ref{fig1}d). Based on well-established literature~\cite{Ross2013,Wang2015a}, the main and lower energy lines are safely assigned to the bright neutral exciton ($\rm X^0$) and to the negative trion ($\rm X^-$), respectively. The negatively charged nature of the trion will be discussed in details below. Next, we also observe spectra that only display emission from the $\rm X^0$ line (middle trace in Fig.~\ref{fig1}d), as well as considerably more complex spectra, showing  a variety of sharper, spatially dependent and spectrally diffusing features assigned to localised excitons~\cite{Branny2016} (bottom trace in Fig.~\ref{fig1}d). These distinct behaviours are directly related to the quality of the MoSe$_2$/FLG and FLG/Au interfaces. In the event of a tight coupling between flat MoS$_2$ and FLG, we expect an efficient ``filtering effect" to take place~\cite{lorchat2020}, which yields single line spectra akin to the middle trace in Fig.~\ref{fig1}d. This effect stems from complete neutralisation of the MoSe$_2$ layer by FLG, together with efficient PL quenching of long-lived localised excitons through non-radiative energy transfer to the FLG, a decay pathway that affects the short-lived $\rm X^0$ emission to a much lesser extent~\cite{lorchat2020}. Hence, the observation of $\rm X^0$ and $\rm X^-$ emission (top trace in Fig.~\ref{fig1}d) suggests partial decoupling between the top MoSe$_2$ layer and the FLG/Au(111) underneath. Dominant emission from localised states (bottom trace in Fig.~\ref{fig1}d) may stem from the local conformation of the MoSe$_2$ layer to the underlying substrate made rougher by thermal annealing. These observations are consistent with complementary $\mu$PL measurements on MoSe$_2$ monolayers directly deposited onto a thermally evaporated polycristalline Au film~\cite{lorchat2020}.

 In such an heterogeneous landscape, STM allows one to probe the topography of a localised area with a resolution down to the atomic scale, making it possible to select regions that are uniform over hundreds of nm$^2$, or individual defects, and eventually address their STML properties. In Fig.~\ref{fig1}e, we show a typical STML spectrum recorded with the STM tip localised on top of an atomically-resolved area of the MoSe$_2$/FLG/Au(111) heterostructure shown in Fig.~\ref{fig1}c. This spectrum is characterised by a prominent emission line located at 1.659 $\pm$ 0.001 eV with a lower-intensity feature at 1.630 $\pm$ 0.001 eV. These two emission lines have full-width at half maximum (FWHM) of 11~meV and 14~meV, respectively. Comparison with the $\mu$PL spectra discussed above allows us to assign the high and low-energy emission lines to $\rm X^0$ and $\rm X^-$, respectively. These spectra provide by far the sharpest excitonic emission lines achieved in STML measurements on TMDs and are the first example where intrinsic luminescence is obtained with the stability and cleanness required for atomically-resolved imaging of the TMD atomic registry. We conclude that the presence of the FLG interlayer preserves the low-temperature luminescence yield of the supported MoSe$_2$ that would otherwise be massively quenched by non-radiative decay channels to the underlying Au(111) substrate~\cite{Krane2016,Pena2020,Velicky2018}.

 The $\rm X^0$ FWHM fitted from our STML spectra, even for the sharpest example (2.9~meV, see Fig.~\ref{fig2}e), still exceeds the homogeneous limit expected in monolayer MoSe$_2$-based vdW heterostructures ($\approx 330~\rm \mu$eV FWHM for a typical  $\rm X^0$ lifetime of 2 ps)~\cite{Fang2019}. This extra broadening may find its origin in the  electronic coupling between TMD and FLG as well as in the excitonic lifetime reduction due to the Purcell effect taking place at the plasmonic tip-sample cavity~\cite{Roslawska2022}.
 
 We estimate an overall quantum yield of $\sim 10^{-7}$ photons/e$^-$ (assuming an overall detection efficiency of $\sim$10$\%$) a value that is orders of magnitude lower than the one reported for single molecules in STML experiments~\cite{Wu2008}. This rather small emission yield suggests that non-radiative exciton decay plays a significant role here. This behaviour may be assigned to the metallic character of graphene that has been shown to efficiently quench hot excitons prior to their relaxation down to the light cone~\cite{lorchat2020,robert2016exciton}. Alternatively, the low electroluminescence yield may reflect an intrinsically weak exciton formation probability, a mechanism that is further discussed below.

STML can be used to identify and characterize inhomogeneities occurring at the scale of  atoms up to a few hundreds of nanometres, and how they affect luminescence properties. To this end, we first recorded a STM image  (\Figref{fig2}a) on a typical area of the heterostructure that approximately corresponds to the area covered by a diffraction-limited laser spot. This image displays flat and well-organised areas separated by ripples, folds and protrusions (arrows in \Figref{fig2}(a)) that stem from the  conformation of the heterostructure to the underlying Au(111) substrate. These protrusions, typically 1~nm-high -- 10~nm-wide, often referred to as nano-bubbles\cite{Darlington2020}, correspond to areas where the TMD and the FLG are slightly decoupled, most likely because of remaining organic adsorbates at the interfaces between MoSe$_2$, FLG and Au(111). These adsorbates segregate locally thanks to the so-called self-cleaning mechanism inherent to 2D materials~\cite{Castellanos_Gomez_2014}. This mechanism that separates clean and flat regions from nano- or microscale pockets of residues is favoured by the annealing step. A pseudo-3D image of one of these protrusions located next to a flat area is provided in \Figref{fig2}b.

STML spectra acquired on top of the protrusion, and $\sim$ 5~nm away on the nearby flat region are displayed in Fig.~\ref{fig2}c. On the flat area, the spectrum (in orange) is characterised by an  $\rm X^0$ line at 1.590 $\pm$ 0.001 eV and by the absence of trion emission, as we shall discuss below. On top of the nano-bubble,the  $\rm X^0$ emission (in blue) is much brighter (by a factor of $\sim 6$) indicating a reduced quenching by the underlying FLG.
Eventually, this spectrum displays an additional peak below the $\rm X^0$ line. The observed redshift (40~meV) is significantly larger than the $\rm X^-$ binding energy (Fig.~1e). This peak is therefore tentatively assigned to excitons localised near defects ($\rm{X^L}$)~\cite{Branny2016}. Similarly, the STML spectrum is strongly altered in the vicinity of larger heterogeneities such as ruptures and folds. An example is provided in the STM image of  Fig.~\ref{fig2}d, where one observes a rupture in the heterostructure (blue cross) next to a flat area (orange cross). 
Corresponding STML spectra are shown in Fig. \ref{fig2}e. In the flat region, the STML spectrum is again characterised by a single, narrow  emission line (2.9~meV FWHM) assigned to $\rm X^0$. 
In contrast, the rupture region displays a complex spectrum, where several narrow resonances below the X$^0$ emission appear, probably arising from localised excitons~\cite{Branny2016}. Interestingly, these lines show widths as narrow as $\sim$ 700 $\mu$eV. Such sharp and red-shifted lines have been reported in PL measurements of defective and highly strained TMDs, where they were attributed to optically active quantum-dots that behave as single photon sources~\cite{He2015,Chakraborty2015,koperski2015single,Srivastava2015,Tonndorf2015,Branny2016,Aharonovich2016,Branny2017,palacios2017}. Time-correlated photon counting measurements~\cite{Merino2015,Zhang2017b}, which are beyond the scope of the present study, would be necessary to ascertain the single-photon emission from these centres. Overall, the data in Fig.~\ref{fig2} indicate that $\mu$PL spectra spatially average emission features from nanoscale regions having distinct spectral responses and emission yields, a complexity that we are able to resolve using our STML approach.

As a next step, we evaluate how the atomic-scale landscape affects excitonic emission from the MoSe$_2$ monolayer. In \Figref{fig3}a-c, we show atomically-resolved STM images of three distinct flat regions of the heterostructure, separated by several micrometres from one another. STML spectra (\Figref{fig3}d) have been acquired in each region for tip positions marked by colored symbols in \Figref{fig3}a-c. In \Figref{fig3}a the STM image reveals the atomic structure of the TMD as well as bright and dark regions indicating smooth height modulation over several nanometres. Here, the STML spectra do not appear to be affected by these modulations, and are characterised by the typical $\rm X^0$ and $\rm X^-$ emission lines. In \Figref{fig3}b and c, we distinguish a moir\'e pattern superimposed on the atomic structure, suggesting a better quality of the MoSe$_2$/FLG interface than in \Figref{fig3}a. This hypothesis is supported by the reduction of the excitonic linewidth from 11 meV (\Figref{fig3}a) to 9 meV (\Figref{fig3}b) and 4 meV (\Figref{fig3}c), respectively. This linewidth narrowing can be assigned to reduced dephasing due to a more homogeneous environment. This higher quality interface may also lead to a more efficient static charge redistribution from the TMD to the FLG flake~\cite{lorchat2020,Hill2017}, explaining the absence of trions in the STML spectrum acquired in \Figref{fig3}b. Noteworthy, within a given nanoscale area, the STML spectra do not depend on the position of the tip with respect to the moir\'e pattern, as revealed in Fig.~\ref{fig3}c.

\begin{figure*}[!th]
  \includegraphics[width=0.65\linewidth]{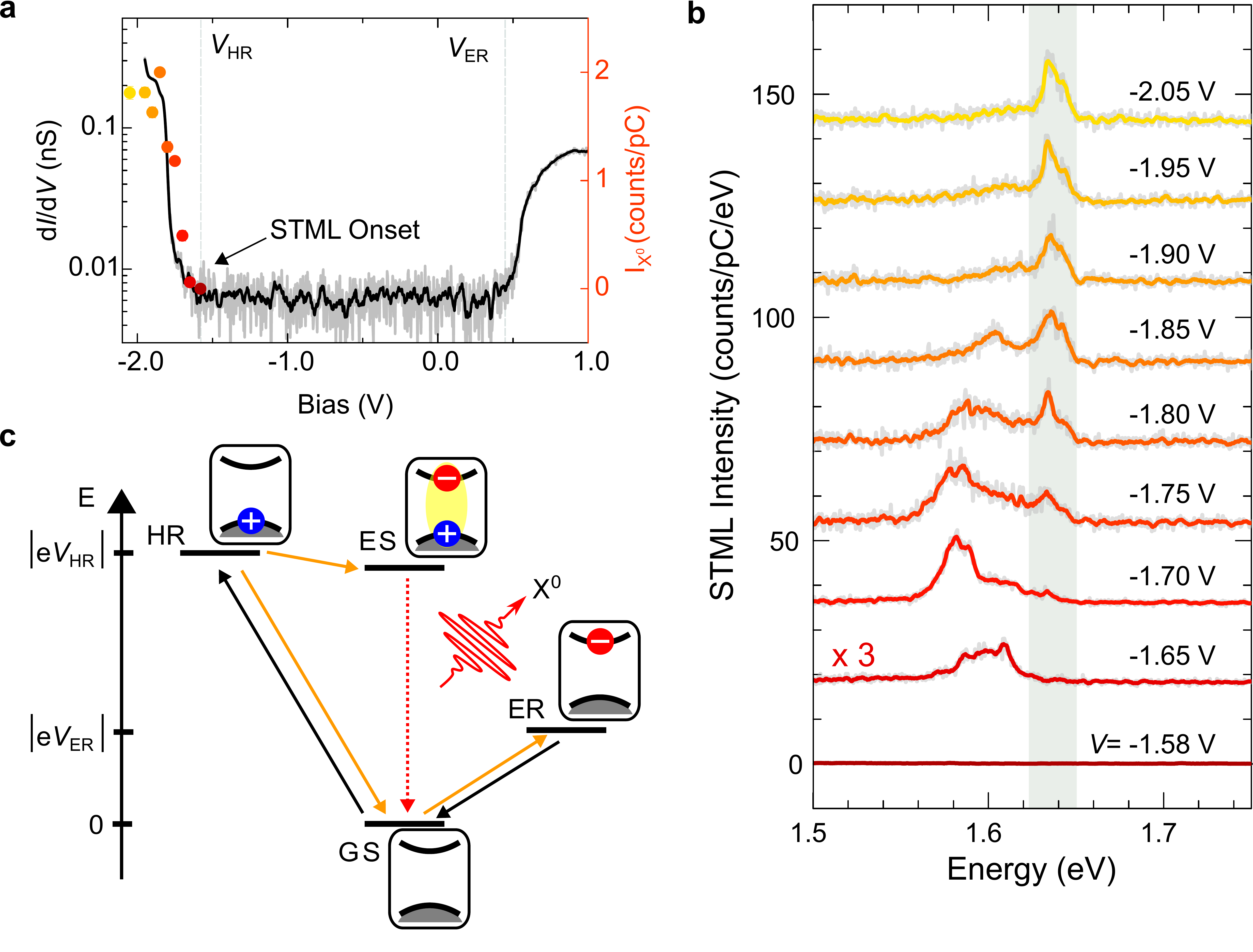}
 \caption{\label{fig4}\textbf{Proposed microscopic mechanism for STML}.~\textbf{a.} Differential conductance spectrum recorded on the MoSe$_2$/FLG surface. The d\textit{I}/d\textit{V} spectrum is overlaid with the integrated intensity associated to the STML-induced X$^0$-emission at different bias voltages (color code identical to the one in \textbf{b}). The close agreement between the two suggests that the luminescence is driven by a charge injection mechanism. \textbf{b.} Bias-dependent STML spectra of the sample recorded under a constant current $I = 30~$pA. The spectra are vertically offset for clarity and the corresponding bias is shown next to each spectra. \textbf{c.} Energy diagram displaying the luminescence mechanism. The system is originally in its ground state (GS) and is driven by a tunneling event into one of two charged resonances: a positively charged resonance denoted HR and a negative one denoted ER. The orange (black) arrows represent an electron (hole) tunneling event. Afterwards, following a second tunneling event, the system relaxes back into GS. This relaxation process can happen either by direct neutralization of HR (ER) or by forming an exciton (denoted ES) which can subsequently recombine radiatively. Note that relaxation through exciton formation is only possible for the HR resonance due to energy conservation.} 
\end{figure*} 

However, in region (c) two emission lines, separated by only 20~meV appear. The most prominent line is strongly redshifted by 70 meV with respect to the $\rm X^0$ line in region (a). Here, an interpretation in terms of defect-induced emission can be ruled out as no atomic defects are imaged in region (c) and we tentatively assign this line to a screened $\rm X^0$. Note that a small $\rm X^0$ redshift of typically 10~meV has already been reported in TMD/monolayer graphene (1LG) and arises from the imperfect cancellation of two effects: screening-induced reduction of the electronic bandgap and of the exciton binding energy ~\cite{lorchat2020,Raja2017}. The unusually large shift of 70~meV observed here could arise from the excellent coupling between the MoSe$_2$/FLG heterostructure and the Au(111) substrate, illustrated by the clear observation of a moir\'e pattern. This tightly coupled stack may favor charge transfer from Au(111) to the MoSe$_2$ layer as well as charge localization assisted by the moir\'e superpotential, hence enabling trion formation. In keeping with this scenario, the faint line located only 20~meV below $\rm X^0$ could arise from a strongly screened trion ($\rm X^-$) with a binding energy reduced by about 10~meV and 5~meV, compared to a bare MoSe$_2$ monolayer~\cite{Ross2013} and to a MoSe$_2$/1LG heterostructure~\cite{lorchat2020}, respectively. In contrast, the absence of emission from trions in region (b) is likely a consequence of the partial decoupling between the coupled MoSe$_2$/FLG heterostructure and the underlying Au(111) substrate, a situation that prevents substrate mediated charge transfer to MoSe$_2$ and trion formation. The distinct signals reported for regions (a), (b) and (c) that enable atomically-resolved STM imaging, therefore hint towards a key role of the MoSe$_2$/FLG and FLG/Au(111) interfaces.

Finally, we jointly address the dependence of the tunnel current and STML spectra with the tip-sample bias $V$ at the same sample spot. The exciton binding energy, $E_{\rm b}$ can in principle be estimated from the difference between the local electronic gap inferred from scanning tunnelling spectroscopy (STS) measurements and the optical gap determined from the $\rm X^0$ energy. Figure ~\ref{fig4}a and b display a differential conductance (${\rm d} I/{\rm d} V$) spectrum and set of STML spectra recorded with increasing bias voltage, respectively. An electronic gap of $2.17 \pm 0.04$ eV is deduced from the ${\rm d} I/{\rm d} V$ spectrum (see Supplementary Section 3 for details), with onsets of the valence and conductance bands at $-1.68 \pm 0.03$~V and $0.49 \pm 0.01$~V respectively, indicating a weak $n$ doping. This doping allows us to identify the trion introduced in \Figref{fig1}c and discussed in Fig.~\ref{fig2} and~\ref{fig3} as negatively charged. The STML spectra in \Figref{fig4}b reveal an $\rm X^0$ emission energy of 1.637~eV, leading to an energy difference of $533 \pm 40$~meV that is close to the value estimated by Ugeda \textit{et al.} on a similar system using a combination of STS and $\mu$PL~\cite{Ugeda2014}. This value is, however, much larger than state-of-the-art optical measurements of $E_{\rm b}$ that converge towards values of 220~meV in hBN-capped monolayer MoSe$_2$~\cite{goryca2019} and  $\approx$ 150~meV in  monolayer MoSe$_2$/1LG heterostructures~\cite{lorchat2020}. Our data therefore demonstrate that the inconsistency between determinations of $E_{\rm b}$ based on the difference between the STS gap and the optical gap versus all-optical measurements does not stem from spatial inhomogenities, as both electronic and optical measurements are local in our approach. We believe that this discrepancy rather originates from the fact that STS measurements poorly resolve electronic states with large in-plane momentum. This phenomenon ultimately leads to an overestimation of the electronic bandgap in STS studies since states near the $\boldsymbol{\Gamma}$ point contribute more to the STS spectrum than those at the edges of the Brillouin zone ($\mathbf{K}$ and $\mathbf{K^\prime}$ points), which define the direct electronic gap in TMD monolayers and from which $\rm X^0$ wavefunctions are formed~\cite{Zhang2015}.

The comparison between STS and STML spectra provides key information regarding the STM-induced exciton formation mechanism. First, no STML signal could be observed under positive $V$. Second, from \Figref{fig4}b, one can deduce an $\rm X^0$ emission onset at $V_0\approx- 1.75~\rm V$. Reporting the integrated intensity of the $\rm X^0$ line as a function of $V$ in the ${\rm d} I/{\rm d} V$ spectrum (color-coded dots in \Figref{fig4}a), one can see that the light emission onset matches well with the onset of the TMD electronic resonance, suggesting that hole injection from the tip is a preliminary step towards excitonic luminescence from the TMD. In \Figref{fig4}c, we propose a simple scheme to explain the luminescence excitation mechanism. This scheme is inspired by a many-body approach  developed to interpret STML data of molecules~\cite{Miwa2019}. Originally in its ground state (GS), the system can be approximated as a two-level system, where the low-energy level is occupied by an electron and the high energy one is empty. GS is here used as the origin of the energy scale on the left of the diagram. At a positive voltage $V_{\rm{ER}} \approx $ 0.49~V, an electron can tunnel from the tip to the TMD (orange arrow) which is driven into a negatively charged (electron) resonance, denoted ER. This state is only transiently populated as the TMD can be efficiently driven back to GS by tunneling of the extra electron to the FLG/Au(111) substrate (black arrow). We can therefore associate the $V = 0.49 $~V feature in the ${\rm d} I/{\rm d} V$ spectrum of \Figref{fig4}a to the GS $\rightarrow$ ER transition. The same reasoning applies for a negative voltage of $V_{\rm{HR}} \approx\, -1.68 ~\rm V$ that corresponds to a transient positively charged (hole) resonance, denoted HR. Here, since $\mid E_{\rm{HR}}\mid >  E_{\rm X^0}$, neutralization by electron transfer from the substrate can lead to the formation of an excited state (ES) in the MoSe$_2$ monolayer that may rapidly relax as an optically active exciton ($X^0$), whose radiative recombination (ES $\rightarrow$ GS, red dotted line) is detected at 1.637~eV. Noteworthy, the $\rm X^0$ STML line is accompanied by lower energy sideband, redshifted by about 40-50~meV and that we assign to STML from $\rm{X^{L}}$. In contrast to $\rm X^0$, the onset for $\rm{X^{L}}$ emission occurs at $V = -1.65 ~\rm V$ (see \Figref{fig4}b). As $\rm X^0$ cannot be generated at this voltage, we deduce that $\rm{X^L}$ are not generated from neutral excitons $\rm X^0$ diffusing to defective areas of the sample, but instead formed directly at the tunneling junction. Their direct excitation may therefore involve in-gap localised electronic states of defects (not identified in the ${\rm d} I/{\rm d} V$ spectrum) or direct energy transfer from inelastic electrons. Further studies are needed to conclude on this point.

In conclusion, we have demonstrated STM-induced excitonic luminescence nanoscopy of an atomically-resolved van der Waals heterostructure. Our work directly reveals how the nanoscale environment influences the  emission characteristics, leading to sizeable excitonic energy shifts and the emergence of emission features from charged and localised excitons on areas separated only by a few nanometres. Hyperspectral mapping of the STML response of atomically-resolved van der Waals heterostructures is still necessary to establish whether STM-induced luminescence imaging reaches sub-nanometre resolution in an extended 2D system, as it does in molecules~\cite{Chen2010,Zhang2016,Doppagne2017,Dole2021,Roslawska2022}, or whether exciton diffusion~\cite{Kulig2018} in TMD monolayers needs to be taken into account.\\
Further STML investigations could enable important advances in quantum materials science. First, STML offers exciting opportunities to unveil near-field charge and energy transfer~\cite{Froehlicher2018,Bradac2021} and proximity effects~\cite{Huang2020,Zutic2019} in vdW heterostructures with unprecedented accuracy, offering natural outcomes in photonics, opto-electronics and nano-electronics. Second, STML is an ideal tool to investigate correlated electronic phases~\cite{Tang2020,Li2021,Wilson2021} and excitons in twist-engineered vdW heterostructures, starting with moir\'e-trapped interlayer excitons~\cite{Seyler2019,Baek2020highly} and trions~\cite{Liu2021} in TMD heterobilayers. Last but not least, STML can be combined with tip-enhanced photoluminescence spectroscopy~\cite{Imada2021} to reach a holistic picture of exciton physics in van der Waals materials at the atomic scale.

\section*{Methods} 

\indent \textbf{Sample fabrication.} The measurements were performed on a vdW heterostructure made using a dry transfer method described in Ref.~\onlinecite{Castellanos_Gomez_2014}. In brief, large-area monolayer MoSe$_2$ and FLG were mechanically exfoliated from their bulk crystals on top of a thin film of polydimethylsiloxane (PDMS). This polymer is used as a temporary substrate in order to fabricate the vdW heterostructure. Its transparency allows for a precise alignment of the layers during the fabrication process. The FLG was then transferred onto a commercial Au(111) thin film grown on mica and the MoSe$_2$ monolayer was afterwards stacked on top of it. The sample was then integrated onto a STM-compatible sample holder, allowing contacting the substrate from two sides (see Supplementary Section 1). The sample was annealed at $200^\circ$C for 24\,h under ultra-high vacuum to reduce surface contamination prior STM measurements.

\indent \textbf{$\boldsymbol{\mu}$PL characterization.} The $\mu$PL response  of our sample was studied in vacuum ($\sim 10^{-5} $ mbar) at a temperature of 14~K in an optical cryostat coupled to a home-built confocal microscopy setup as in Ref.~\onlinecite{lorchat2020}. A linearly polarised continuous-wave laser beam at a wavelength of 532~nm (2.33~eV) with an intensity of 30~$\mu \rm{W}/\mu \rm{m}^2$ at the sample was used for all measurements. The $\mu$PL signal was collected in a backscattering geometry and dispersed onto a liquid-nitrogen cooled CCD array using a 500~mm monochromator equipped with a 150 grooves/mm grating.

\indent \textbf{STM and STML measurements.} STM-based measurements were performed in a low temperature (6\,K) Unisoku STM operating in ultrahigh vacuum and adapted to optical measurements. A first lens (numerical aperture of 0.55), mounted on a three axis piezo controller, is used to collect and collimate the light emitted at the tip-sample junction. This light is then redirected outside of the vacuum chamber through successive windows and viewports. Afterwards, it is refocused on an optical fiber, itself connected to a spectrometer, and eventually detected using a nitrogen-cooled CCD camera. Three different gratings were used, yielding spectral resolutions ranging from 0.6~nm down to 0.06~nm. Silver (Ag) STM tips were used to optimise the plasmonic response of the junction and tested on a separate Ag(111) sample to maximise the radiative recombination rate at the tip-sample junction. The STS spectrum shown in Fig. \ref{fig4}a was acquired at constant height with a current setpoint $I = 30~$pA and a modulation voltage $V_{mod} = 20~$mV.

\indent \textbf{Data analysis} All $\mu$PL and STML spectra presented in this work present both raw (light gray) and the smoothed data (solid colored line). Unless specified in the text, all spectra were fitted using Voigt profiles.

~

\section*{Acknowledgments}
We thank Michael Chong, Benjamin Doppagne, Guillaume Froehlicher, Arnaud Gloppe, Eric Le Moal, Etienne Lorchat and Tom\'a\v{s} Neuman for fruitful discussions. We are grateful to the IPCMS mechanical workshop, in particular Halit Sumar, and Virginie Speisser, Michelangelo Romeo and the STnano clean room staff for technical support.  This project has received funding from the European Research Council (ERC) under the European Union's Horizon 2020 research and innovation program (grant agreement No 771850) and the European Union's Horizon 2020 research and innovation program under the Marie Sk\l{o}dowska-Curie grant agreement No 894434. We acknowledge financial support from the Agence Nationale de la Recherche under grant ATOEMS ANR-20-CE24-0010. This work of the Interdisciplinary Thematic Institute QMat, as part of the ITI 2021 2028 program of the University of Strasbourg, CNRS and Inserm, was supported by IdEx Unistra (ANR 10 IDEX 0002), and by SFRI STRAT'US project (ANR 20 SFRI 0012) and EUR QMAT ANR-17-EURE-0024 under the framework of the French Investments for the Future Program.  S.B. acknowledges support from the Indo-French Centre for the Promotion of Advanced Research (CEFIPRA) and from the Institut Universitaire de France (IUF).

%


\onecolumngrid
\newpage
\begin{center}
{\Large\textbf{Supplementary Information for: \\ Tip-induced excitonic luminescence nanoscopy of an atomically-resolved van der Waals heterostructure}}
\end{center}

\setcounter{equation}{0}
\setcounter{figure}{0}
\setcounter{section}{0}
\renewcommand{\thetable}{S\arabic{table}}
\renewcommand{\theequation}{S\arabic{equation}}
\renewcommand{\thefigure}{S\arabic{figure}}
\renewcommand{\thesection}{S\arabic{section}}
\renewcommand{\thesubsection}{S\arabic{section}\alph{subsection}}
\renewcommand{\thesubsubsection}{S\arabic{section}\alph{subsection}\arabic{subsubsection}}

\linespread{1.4}\selectfont




\section{Sample architecture}
\begin{figure}[h!]
\centering
\includegraphics[scale=0.4]{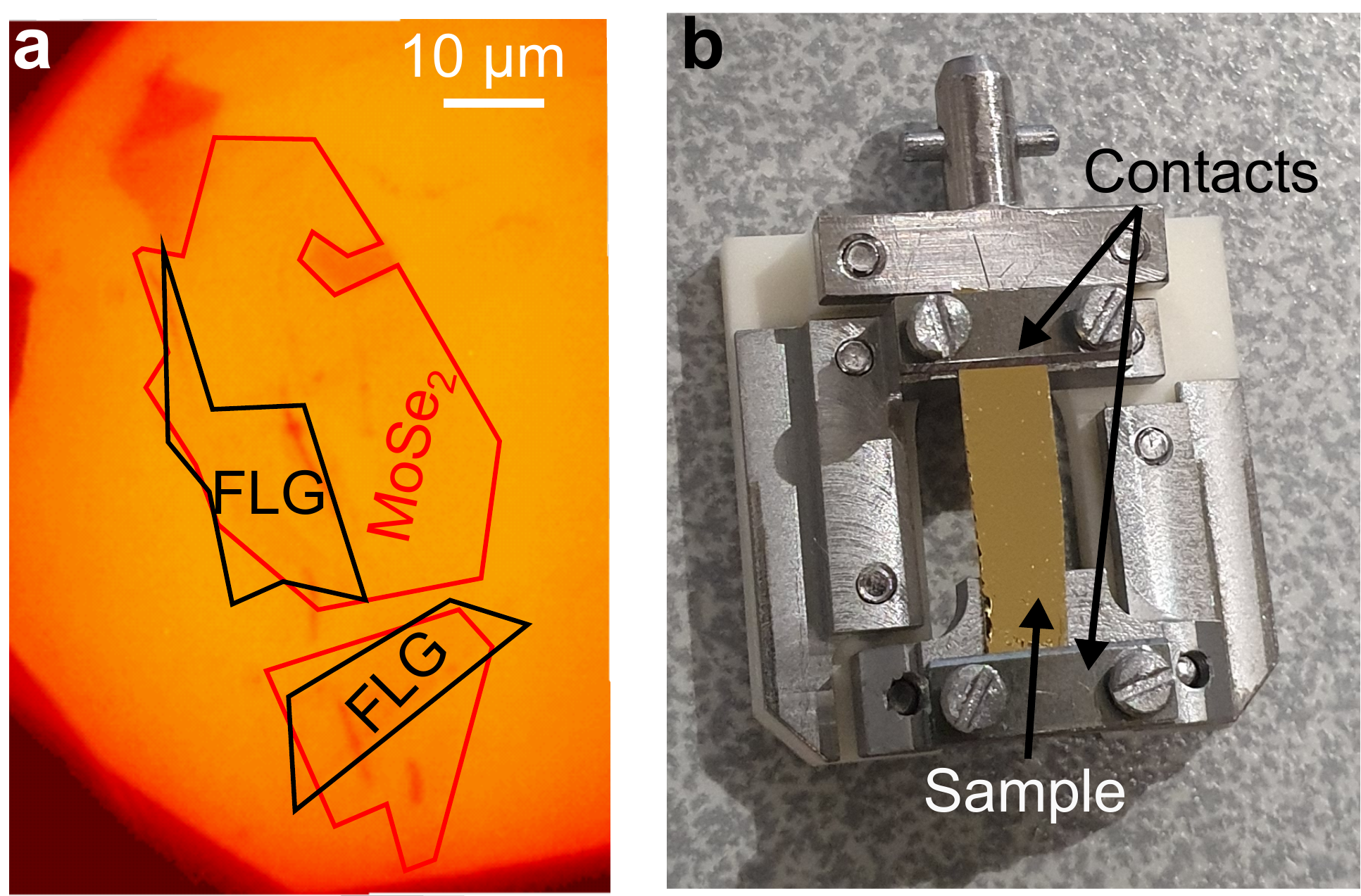}
\caption{\textbf{Sample design}. \textbf{a}. Optical microscopy image of a MoSe$_2$/few-layer graphene (FLG) heterostructure deposited onto an Au(111) film on mica. \textbf{b}. Picture of the sample holder.}
\label{FigSI_1}
\end{figure}

The sample consists of a monolayer of MoSe$_2$ stacked on top of a few ($\sim 3-5$) layer graphene flake (FLG). The layers were isolated by mechanical exfoliation from bulk crystals and subsequent deposition onto polydimethylsiloxane (PDMS) films. After optical identification of individual layers, the stacking was performed in a homemade stamping station following the dry transfer method described in Ref.~\cite{Castellanos_Gomez_2014}.To ensure compatibility with our STM setup, we used, as a substrate, a commercial thin Au(111) film grown on mica. An optical image of the sample is shown in Fig.~\ref{FigSI_1}a. Finally, the sample was integrated into a STM-compatible sample holder, which allows us to contact the substrate from two sides. The final configuration of the sample is shown in Fig.~\ref{FigSI_1}b.

\par

\newpage

\section{Micro-photoluminescence ($\mu$PL) characterization at 14 K}

Following the fabrication, the sample is rapidly introduced in a preparation chamber and annealed for 24~h at 200~$^{\circ}\text{C}$ under ultra high vacuum (UHV) conditions. Although this step is crucial to perform STM measurements, it can significantly modify the optical properties of the monolayer \cite{lorchat2020,Pena2020}. The rise in temperature enhances contaminant diffusion at the different interfaces, resulting in a spatially-dependent coupling between the component layers and between the heterostructure and the substrate itself. These variations would then modify nanoscale landscape and ultimately determine the optical properties of the heterostructure. At room temperature, it has been observed that the annealing of mechanically exfoliated TMD on Au(111) can lead to a drastic quenching of the PL and STML signals~\cite{Pena2020}. At cryogenic temperatures, a wide variety of spectra have been reported in TMD/Au heterostructures. From "typical" spectra akin to the ones observed on monolayer TMD deposited onto non-metallic substrates, to spectra composed of multiple resonances and dominated by dielectric screening and quenching~\cite{Grzeszczyk2020,lorchat2020}. In the case of extremely clean TMD/metal interfaces, such as the ones obtained using epitaxial growth, a complete quenching of the luminescence has been reported~\cite{Krane2016,Schuler2020}. This points towards the interface quality as the main driver of the optical properties of the heterostructure.

\par

\begin{figure}[h!]
\centering
\includegraphics[scale=0.28]{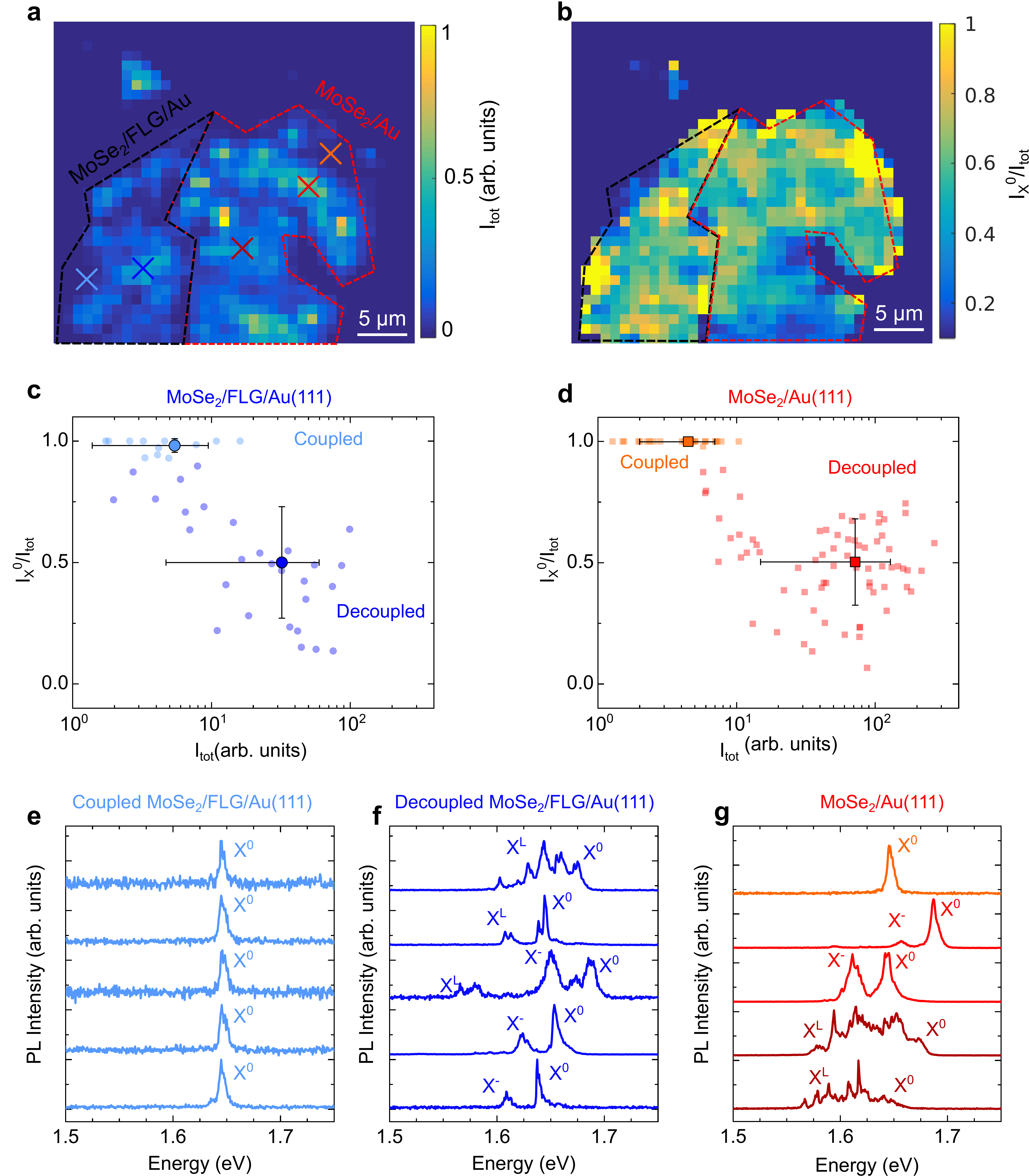}
\caption{\textbf{Low temperature $\boldsymbol{\mu}$PL mapping.} Hyperspectral $\mu$PL maps of the MoSe$_2$/FLG/Au(111) sample studied in the main manuscript showing \textbf{a} the total integrated intensity (denoted $\rm{I}_{\rm{tot}}$) and \textbf{b} the ratio between the integrated intensity from the exciton emission ($\rm{I}_{\rm{X}^0}$) and the total integrated $\mu$PL intensity. Correlations between $\rm{I}_{\rm{X}^0}/\rm{I}_{\rm{tot}}$ and $\rm{I}_{tot}$ for \textbf{c} MoSe$_2$/FLG/Au(111) and  \textbf{d} MoSe$_2$/Au(111) \textbf{e}. Series of $\mu$PL spectra acquired at the positions marked in \textbf{a}. The spectra are representative of coupled MoSe$_2$/FLG/Au(111) area \textbf{e}, decoupled MoSe$_2$/FLG/Au(111) \textbf{f} and MoSe$_2$/Au(111) \textbf{g}. All spectra were acquired at 14 K under continuous wave laser-excitation centered at 532 nm and under an incident laser intensity of about $\sim 30 ~ \mu$W/$\mu\rm{m}^2$.}
\label{Fig2_SI}
\end{figure}

In our case, the self-cleaning mechanism clusters contaminants together at the MoSe$_2$/FLG interface. This ensures the presence of locally contaminant-free areas as can be seen in Fig.~2a in the main text. As a result, the nanoscopic landscape of the sample is composed of different ``patches" of homogeneous areas displaying varying degrees of coupling, strain and screening. This heterogeneity has a direct impact on the $\mu$PL spectra, which spatially average excitonic recombination over a diffraction-limited area. Figure \ref{Fig2_SI}a shows a hyperspectral map of the total integrated PL intensity ($\rm{I}_{\rm{tot}}$) of the MoSe$_2$/FLG/Au(111) heterostructure. We observe $\mu$PL signal variations of one order of magnitude between coupled (dim areas) and decoupled areas (bright areas) in both  MoSe$_2$/FLG/Au(111) and MoSe$_2$/Au(111) surfaces. This is a direct signature of the local interface quality between the components. In the case of MoSe$_2$/FLG, there are two interfaces that may play a role: the MoSe$_2$/FLG interface regulating exciton transfer from MoSe$_2$ to graphene and the FLG/Au(111) interface, which can further contribute to luminescence quenching and dielectric screening. We can illustrate this by plotting, in Fig.~\ref{Fig2_SI}b, the ratio between the integrated intensity of the neutral exciton ($\rm{X}^0$) emission ($\rm{I}_{\rm{X}^0}$) and $\rm{I}_{\rm{tot}}$. Qualitatively, We observe that the dimmest areas (low $I_{\rm{tot}}$) tend to display near unity $I_{\rm X^0}/I_{\rm{tot}}$ ratios. To substantiate this observation, we have plotted the correlation between the observed ratio and $\rm{I}_{\rm{tot}}$ for spectra taken from coupled and decoupled areas of the heterostructure (Fig.~\ref{Fig2_SI}c-d). In both regions, MoSe$_2$/FLG/Au(111) and MoSe$_2$/Au(111), a correlation between large $I_{\rm X^0}/I_{\rm{tot}}$ ratios and correspondingly low $\rm{I}_{\rm{tot}}$ is revealed. This is consistent with a scenario where the layers that compose the heterostructure are well coupled and the emission arises solely from $\rm X^0$, as observed in samples encapsulated in hexagonal Boron Nitride (hBN)~\cite{lorchat2020}. In inhomogeneous areas, the presence of contaminants and bubbles contributes to exciton localization, resulting in extrinsic, low energy features in the $\mu$PL spectra, which reduce the overall ratio.

Typical $\mu$PL spectra recorded on the positions marked in Fig.~\ref{Fig2_SI}a are shown in Fig.~\ref{Fig2_SI}e-g. The spectra are characteristic of the main areas of the sample (also summarised in Fig.~1d: coupled MoSe$_2$/FLG/Au(111) (light blue), decoupled MoSe$_2$/FLG/Au(111) (blue), coupled MoSe$_2$/Au(111) (orange), decoupled MoSe$_2$/Au(111) (red) and MoSe$_2$/Au(111) in the vicinity of bubbles (dark red). In coupled areas (Fig.~\ref{Fig2_SI}e and g), single-line $\mu$PL spectra are observed, as previously reported in hBN-capped TMD/graphene heterostructures~\cite{lorchat2020}. On the other hand, decoupled areas display multiple low-energy peaks arising from neutral, charged and localised excitons ($\mathrm X^0$, $\mathrm X^-$ and $\mathrm {X^L}$ respectively) as shown in Fig.~\ref{Fig2_SI}f,g. The number of peaks and their positions are spatially dependent and are a direct signature of the homogeneity of the probed area. Nevertheless, one can recover "typical" spectra composed of two peaks ($\mathrm X^0$ and $\mathrm X^-$). Interestingly, we observe redshifts of the excitonic emission as large as 40~meV within decoupled regions (bottom spectra in Fig.~\ref{Fig2_SI}f and red spectra in Fig \ref{Fig2_SI}g). We speculate that these redshifts are due to a combination of strain and screening from the Au(111) substrate.
\par

\newpage

\section{Determination of the electronic gap}

To determine the onset of the bands in the differential conductance spectra (see Fig.~4a in the main text), we used a similar procedure as in Ref.~\onlinecite{Ugeda2014}. This procedure is illustrated in Fig.~\ref{thegap} and can be summarised as follows:

\begin{figure}[ht!]
\centering
\includegraphics[scale=0.77]{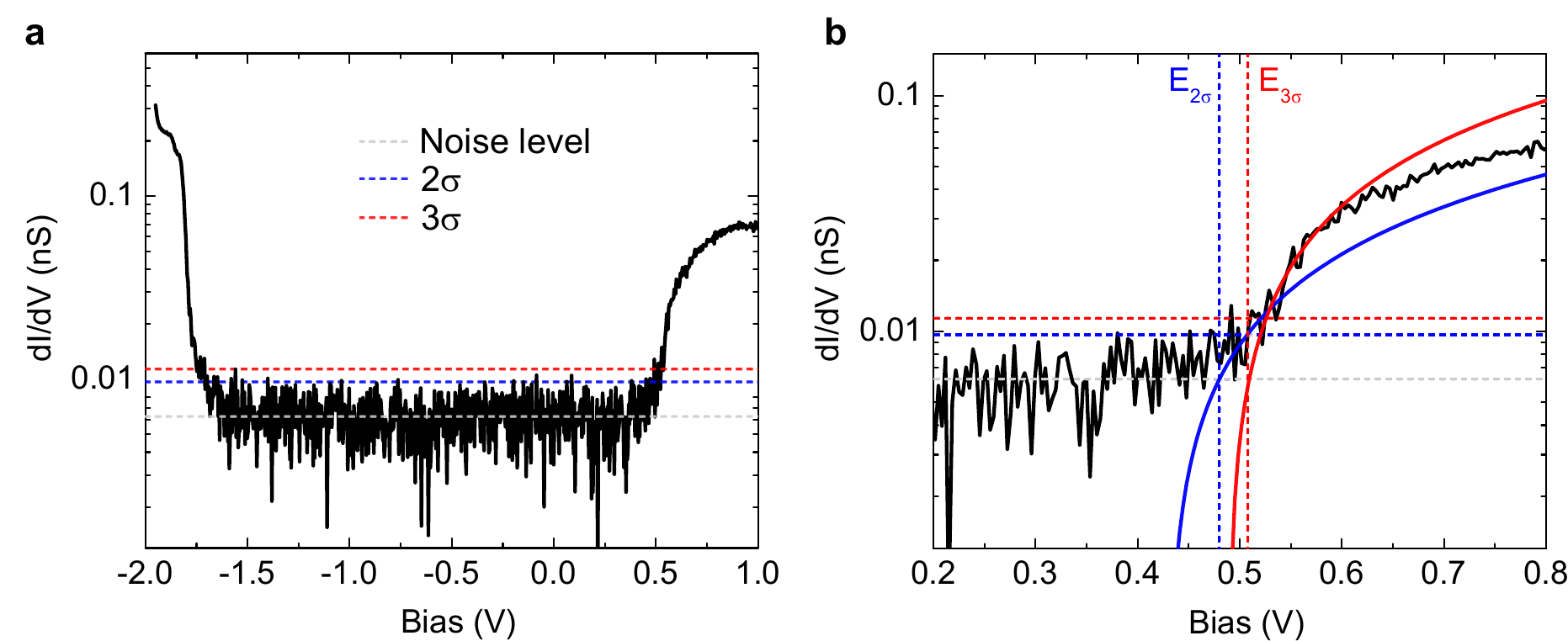}
\caption{\textbf{Determination of the band onsets and electronic gap.} \textbf{a.} Differential conductance spectrum shown in Fig.~4a of the main manuscript. The spectrum is shown in semi-logarithmic scale and it has been vertically offset to avoid negative values. The gray dashed line corresponds to the average noise level. The blue (red) horizontal dashed line marks a threshold for which the signal is $2\sigma$ ($3\sigma$) above the noise level.~\textbf{b.} Differential conductance spectrum shown in \textbf{a} around the negative resonance onset. The value of the onset is determined by fitting the band at a given threshold value. The linear fits are represented by solid lines and correspond to a $2\sigma$ (blue) and $3\sigma$ (red) threshold.}
\label{thegap}
\end{figure}

\vspace{-0.5 cm}

\begin{itemize}
    \item First, we plot the spectrum on a semi-logarithmic scale (Fig.~\ref{thegap}a). A rigid vertical offset was applied to avoid running into negative values.
    \item To estimate the noise level of the measured conductance, we compute the average value (gray dashed line in Fig.~ \ref{thegap}a) and the standard deviation of the noise inside the gap.
    \item We defined two thresholds from which we consider that the signal is sufficiently different from the noise floor. These correspond to $2 \sigma$ and $3\sigma$  ($\sigma$ being the standard deviation) above the average noise. These thresholds are represented by the blue and red dashed lines in Fig.~\ref{thegap}a, respectively.   
    \item Finally, we used linear fits of the bands within a range of $\mathrm E_{\mathrm{VB},n\sigma}-\Delta \mathrm E < \mathrm E <\mathrm E_{\mathrm{VB},n\sigma}$ and $\mathrm E_{\mathrm{CB},n\sigma} <\mathrm E <\mathrm E_{\mathrm{CB},n\sigma}+\Delta \mathrm E$ with $n = 2,~3$. $\Delta \mathrm E$ was chosen so that the Pearson coefficient $R^2$ of the fits is not smaller than 0.95.
    \item The onset of the bands is then defined as the intersection between the linear fit and the floor level for each threshold. We then calculate the average of both values. 
    \item These values are marked with the dashed vertical lines in Fig.~\ref{thegap}b.
\end{itemize}
  
Using this procedure, we could determine an electronic gap of $2.17 \pm 0.04 $~eV all over the sampled region.


\end{document}